\documentclass[fleqn,twoside]{article}
\usepackage{espcrc2}

\newcommand{\ltwid}{\mathrel{\raise.3ex\hbox{$<$\kern-.75em\lower1ex\hbox{$\sim$}}}}
\newcommand{\AmS}{{\protect\the\textfont2
  A\kern-.1667em\lower.5ex\hbox{M}\kern-.125emS}}

\title{A Leading Log Approximation for Inflationary Quantum Field Theory}

\author{R. P. Woodard\address{Department of Physics, University of 
        Florida, Gainesville, FL 32611, United States}
        \thanks{woodard@phys.ufl.edu}}
       
\begin{document}

\begin{abstract}
During inflation explicit perturbative computations of quantum 
field theories which contain massless, non-conformal fields 
exhibit secular effects that grow as powers of the logarithm of 
the inflationary scale factor. Starobinski\u{\i}'s technique of
stochastic inflation not only reproduces the leading infrared
logarithms at each order in perturbation theory, it can sometimes
be summed to reveal what happens when inflation has proceeded so
long that the large logarithms overwhelm even very small coupling
constants. It is thus a cosmological analogue of what the 
renormalization group does for the ultraviolet logarithms of 
quantum field theory, and generalizing this technique to quantum 
gravity is a problem of great importance. There are two significant
differences between gravity and the scalar models for which 
stochastic formulations have so far been given: derivative 
interactions and the presence of constrained fields. We use 
explicit perturbative computations in two simple scalar models 
to infer a set of rules for stochastically formulating theories 
with these features.
\vspace{1pc}
\end{abstract}

\maketitle

\section{Overview}

This conference report represents work done in collaboration with
Nikolaos Tsamis of the University of Crete. In the next section I
review the two conditions (masslessness and lack of conformal 
invariance) for quantum effects to be enhanced during inflation. 
In section 3 I introduce the notions of infrared logarithms and
the leading logarithm approximation. Section 4 shows that the
field operator behaves like a classical random variable in the
leading logarithm approximation. I also derive the approximate
field equation obtained by Starobinski\u{\i} for a massless, 
minimally coupled scalar with non-derivative self-interactions.
Sections 5 and 6 generalize Starobinski\u{\i}'s technique to 
theories which possess derivative interactions and constraints,
respectively. Our results are summarized in section 7.

\section{Quantum Effects during Inflation}

A locally de Sitter geometry provides the simplest paradigm for inflation.
To see why, consider a general homogeneous, isotropic and spatially flat 
geometry,
\begin{equation}
ds^2 = -dt^2 + a^2(t) d\vec{x} \cdot d\vec{x} \; . \label{dS}
\end{equation}
Derivatives of the scale factor $a(t)$ give the Hubble parameter 
$H(t)$ and the deceleration parameter $q(t)$,
\begin{equation}
H(t) \equiv \frac{\dot{a}}{a} \qquad , \qquad q(t) \equiv 
-\frac{a \ddot{a}}{\dot{a}^2} = -1 - \frac{\dot{H}}{H^2} \; . \label{Hq}
\end{equation}
The nonzero components of the Riemann tensor are,
\begin{eqnarray}
R^0_{~i0j} & = & - q H^2 g_{ij} \; , \nonumber \\
R^i_{~jk\ell} & = & H^2 \Bigl(\delta^i_{~k} g_{j\ell} - \delta^i_{~\ell} 
g_{jk}\Bigr) \; . \label{Riemann}
\end{eqnarray}
Inflation is defined as positive expansion ($H(t) > 0$) with 
negative deceleration ($q(t) < 0$). On the other hand, stability --- in
the form of the weak energy condition --- implies $q(t) \geq -1$. At the
limit of $q = -1$ we see from (\ref{Riemann}) that the Riemann tensor 
assumes the locally de Sitter form, $R^{\rho}_{~\sigma \mu \nu} = H^2 
\Bigl(\delta^{\rho}_{~\mu} g_{\sigma \nu} - \delta^{\rho}_{~\nu} 
g_{\sigma \mu} \Bigr)$. It follows from (\ref{Hq}) that the Hubble 
parameter is actually constant, and that the zero of time can be chosen 
to make the scale factor take the simple exponential form we shall 
henceforth assume,
\begin{equation}
{\rm de\ Sitter\ inflation} \qquad \Longrightarrow \qquad a(t) = e^{Ht} \; .
\end{equation}

The homogeneity of spacetime expansion in (\ref{dS}) does not change the
fact that particles have constant wave vectors $\vec{k}$, but it does 
alter what these means physically. In particular the energy of a particle
with mass $m$ and wave number $k = \Vert \vec{k}\Vert$ becomes time dependent,
\begin{equation}
E(t,k) = \sqrt{m^2 + \Bigl(k/a(t)\Bigr)^2} \; .
\end{equation}
This results in an interesting change in the en\-er\-gy-time uncertainty
principle which restricts how long a virtual pair of such particles
with $\pm \vec{k}$ can exist. If the pair was created at time $t$ it
can last a time $\Delta t$ given by the integral,
\begin{equation}
\int_t^{t+\Delta t} \!\!\!\!\! dt' E(t',k) \sim 1 \; .
\end{equation}
Just as in flat space, particles with the smallest masses persist 
longest. For the fully massless case the integral is simple to evaluate,
\begin{equation}
\int_t^{t+\Delta t} \!\!\!\!\! dt' E(t',k) \Bigl\vert_{m=0} = 
\Bigl[1 - e^{-H \Delta t}\Bigr] \frac{k}{H a(t)} \; .
\end{equation}
We therefore conclude that any massless virtual particle which happens 
to emerge from the vacuum with $k \ltwid H a(t)$ can persist forever!

Most massless particles possess conformal invariance. The change of
variables $d\eta \equiv dt/a(t)$ defines a conformal time in terms of
which the invariant element (\ref{dS}) is just a conformal factor 
times that of flat space,
\begin{equation}
ds^2 = -dt^2 + a^2(t) d\vec{x} \cdot d\vec{x} = a^2 \Bigl(-d\eta^2
+ d\vec{x} \cdot d\vec{x}\Bigr) \; . 
\end{equation}
In the $(\eta,\vec{x})$ coordinates conformally invariant theories 
are locally identical to their flat space cousins. The rate at which 
virtual particles emerge from the vacuum per unit conformal time must 
be the same constant --- call it $\Gamma$ --- as in flat space. Hence 
the rate of emergence per unit physical time is,
\begin{equation}
\frac{dN}{dt} = \frac{dN}{d\eta} \frac{d\eta}{dt} = \frac{\Gamma}{a(t)} \; .
\end{equation}
It follows that, although any sufficiently long wavelength, massless 
and conformally invariant particle which emerges from the vacuum can
persist forever during inflation, very few will emerge.

Two kinds of massless particles do not possess conformal invariance:
minimally coupled scalars and gravitons. To see that the production of
these particles is not suppressed during inflation note that each 
polarization and wave number behaves like a harmonic oscillator with 
time dependent mass and frequency,
\begin{eqnarray}
\lefteqn{L = \frac12 m \dot{q}^2 - \frac12 m \omega^2 q^2} \nonumber \\
& & {\rm with} \quad m(t) = a^3(t) \quad {\rm and} \quad 
\omega(t) = \frac{k}{a(t)} \; .
\end{eqnarray}
The Heisenberg equation of motion can be exactly solved,
\begin{eqnarray}
\lefteqn{\ddot{q} + 3 H \dot{q} + \frac{k^2}{a^2} q = 0} \nonumber \\
& & \Longrightarrow \qquad q(t) = u(t,k) \alpha + u^*(t,k) \alpha^{\dagger} \; ,
\end{eqnarray}
where the mode functions and commutation relations are,
\begin{eqnarray}
\lefteqn{u(t,k) = \frac{H}{\sqrt{2 k^3}} \Biggl[1 - \frac{i k}{H a(t)}\Biggr] 
\exp\Biggl[\frac{ik}{Ha(t)}\Biggr] } \nonumber \\
& & {\rm and} \qquad [\alpha,\alpha^{\dagger}] = 1 \; .
\end{eqnarray}
The (co-moving) energy operator for this system is,
\begin{equation}
E(t) = \frac12 m(t) \dot{q}^2(t) + \frac12 m(t) \omega^2(t) q^2(t) \; .
\end{equation}
Owing to the time dependent mass and frequency, there are no stationary 
states for this system. At any given time the minimum eigenstate of
$E(t)$ has energy $\frac12 \omega(t)$, but which state this is changes 
for each value of time. The state $\vert \Omega \rangle$ which is 
annihilated by $\alpha$ has minimum energy in the distant past. The 
expectation value of the energy operator in this state is,
\begin{eqnarray}
\lefteqn{\Bigl\langle \Omega \Bigl\vert E(t) \Bigr\vert \Omega \Bigr\rangle =
\frac12 a^3(t) \vert \dot{u}(t,k) \vert^2 + \frac12 a(t) k^2
\vert u(t,k) \vert^2 } \nonumber \\
& & = \frac{k}{2a} + \frac{H^2 a}{4k} \; .
\end{eqnarray}
If we think of each particle having energy $k/a(t)$ it follows that the
number of particles with any polarization and wave number $k$ grows as
the square of the inflationary scale factor,
\begin{equation}
N(t,k) = \Biggl(\frac{H a(t)}{2 k}\Biggr)^2 \; ! \label{Ngrowth}
\end{equation}

Quantum field theoretic effects are driven by essentially classical 
physics operating in response to the source of virtual particles 
implied by quantization. On the basis of (\ref{Ngrowth}) one might 
expect inflation to dramatically enhance quantum effects from massless, 
minimally coupled scalars and gravitons, and explicit studies over a
quarter century have confirmed this. The oldest results are of course 
the cosmological perturbations induced by scalar inflatons \cite{MC} 
and by gravitons \cite{AAS1}. More recently it was shown that the one 
loop vacuum polarization induced by a charged, massless and minimally 
coupled scalar causes super-horizon photons to behave like massive 
particles in some ways \cite{PTW1,PTW2,PW1}. Another recent result is 
that the one loop fermion self-energy induced by a massless and minimally 
coupled Yukawa scalar reflects so much particle production that the end 
of inflation should reveal a degenerate gas of super-horizon fermions 
\cite{PW2}. 

\section{Infrared Logarithms and the Leading Log Approximation}

Two higher loop results are of special interest to us. The first is 
that the gravitational back-reaction from the inflationary production 
of gravitons induces an ever greater slowing in the expansion rate
\cite{TW1,TW2,TW3}. The Lagrangian is just that of general relativity 
with a positive cosmological constant,
\begin{equation}
\mathcal{L} = \frac1{16 \pi G} \Bigl(R - 2 \Lambda\Bigr) \sqrt{-g} 
\quad {\rm with} \quad \Lambda = 3 H^2 \; . \label{Lgr}
\end{equation}
What was actually computed \cite{TW3} is the graviton one-point function,
about a locally de Sitter background, in the presence of a state which 
is free Bunch-Davies vacuum at $t=0$. However, if the resulting distortion
of the background geometry was viewed in terms of an effective energy 
density and pressure the results would be,
\begin{eqnarray}
\lefteqn{\rho(t) = \frac{\Lambda}{8 \pi G} + \frac{(\kappa H)^2 H^4}{2^6 \pi^4}
} \nonumber \\
& & \hspace{.4cm} \times \Biggl\{-\frac12 \ln^2(a) + O\Bigl(\ln(a)\Bigr) 
\Biggr\} + O(\kappa^4) , \label{rhoGR} \\
\lefteqn{p(t) = -\frac{\Lambda}{8 \pi G} + \frac{(\kappa H)^2 H^4}{2^6 \pi^4}}
\nonumber \\
& & \hspace{.4cm} \times \Biggl\{+\frac12 \ln^2(a) + O\Bigl(\ln(a)\Bigr) 
\Biggr\} + O(\kappa^4) . \label{pGR}
\end{eqnarray}
(Here $\kappa^2 \equiv 16 \pi G$.) Of course the tree order ($\sim 
\kappa^{-2}$) term is just the cosmological constant. The one loop 
($\sim \kappa^0$) effect from the kinetic energies of inflationary
gravitons is a constant \cite{Ford,FMVV} that must be subsumed into 
$\Lambda$ in order for the universe to begin inflating with Hubble
parameter $H$. The next order effect is secular because the kinetic 
energies interact with the total graviton field strength which grows 
as more and more gravitons are produced. The induced energy density 
of this term is negative because the interaction is attractive.

The second result is that the inflationary particle production of
self-interacting scalars induces a violation of the weak energy condition
on cosmological scales \cite{OW1,OW2}. The Lagrangian is that of a 
massless, minimally coupled scalar with a quartic self-interaction in 
a non-dynamical, locally de Sitter background,
\begin{equation}
\mathcal{L} = -\frac12 \partial_{\mu} \varphi \partial_{\nu} \varphi
g^{\mu\nu} \sqrt{-g} - \frac{\lambda}{4!} \varphi^4 \sqrt{-g} 
\; . \label{Lphi}
\end{equation}
This theory can be renormalized so that, when released in free 
Bunch-Davies vacuum at $t=0$, the energy density and pressure are 
\cite{OW1,OW2},
\begin{eqnarray}
\lefteqn{\rho(t) = \frac{\Lambda}{8 \pi G} + 
\frac{\lambda H^4}{2^6 \pi^4} \Biggl\{\frac12 \ln^2(a) } \nonumber \\
& & \hspace{1cm} + \Delta \xi \ln(a) + O\Bigl(a^{-1} \Bigr) \Biggr\} + 
O(\lambda^2) , \label{rhophi} \\
\lefteqn{p(t) = -\frac{\Lambda}{8 \pi G} + \frac{\lambda H^4}{2^6 \pi^4}
\Biggl\{-\frac12 \ln^2(a) } \nonumber \\
& & - \Bigl[\frac13 \!+\! \Delta \xi\Bigr] \ln(a) + O\Bigl(a^{-1}\Bigr) 
\Biggr\} + O(\lambda^2) . \label{pphi}
\end{eqnarray}
(The parameter $\Delta \xi$ is related to the finite part of the 
conformal counterterm.) The one loop effect from the kinetic 
energies of inflationary scalars is again a constant that must be 
subsumed into $\Lambda$. The next order effect is secular because 
the $\varphi^4$ self-interaction involves the total scalar field 
strength which grows as more and more scalars are produced. The 
induced energy density is positive, for $\lambda > 0$, because the 
$\varphi^4$ term adds to the energy density. In this model we might 
also guess that the effect is self-limiting because the classical 
restoring force tends to push the scalar back to zero.

Both of these models exhibit {\it infrared logarithms} --- factors of
$\ln(a) = Ht$. As inflation proceeds these infrared logarithms grow
without bound until they eventually overcome the small coupling 
constants, $(\kappa H)^2$ for gravity and $\lambda$ for the scalar model.
One cannot conclude from this that there is actually a significant
change in the background expansion rate because we do not know the
higher order results. The legitimate conclusions are rather (1) that 
the expansion rate slows in the first model, and increases in the 
second, and (2) that both effects eventually become nonperturbatively 
strong. To learn what the outcome is at late times, after perturbation 
theory has broken down, requires a nonperturbative technique.

One possibility for such a technique is to sum the leading infrared
logarithms. It has been conjectured that the powers of $a^{-1} = 
e^{-Ht}$ --- which are separately conserved --- can be subsumed into 
renormalizations of the initial state \cite{OW2}. If this is so then we 
expect the perturbative expansions for the energy density in both models 
to take the general form,
\begin{eqnarray}
\lefteqn{\rho(t) = \frac{\Lambda}{8 \pi G} + H^4 \sum_{\ell = 2}^{\infty} 
\Bigl({\rm coupling}\Bigr)^{\ell-1} } \nonumber \\
& & \hspace{1.5cm} \times \sum_{k=0}^{2\ell-3} C_{\ell k} \Bigl(\ln(a)
\Bigr)^{2 \ell -2 - k} ,
\end{eqnarray}
where the $C_{\ell k}$'s are pure numbers, and ``coupling'' is 
$(\kappa H)^2$ for gravity and $\lambda$ for the scalar model. The leading 
logarithm approximation would be,
\begin{eqnarray}
\lefteqn{\rho(t)_{{\rm leading} \atop {\rm logarithm}} = \frac{\Lambda}{8 
\pi G} } \nonumber \\
& & \hspace{.8cm} + H^4 \sum_{\ell = 2}^{\infty} C_{\ell 0} 
\Bigl({\rm coupling} \; \ln^2(a) \Bigr)^{\ell-1} .
\end{eqnarray}

Of course other quantities besides the energy density exhibit infrared 
logarithms. For example, they have been evaluated at the one \cite{OW1} 
and two loop \cite{TOW} orders in the self-mass-squared of the scalar 
model (\ref{Lphi}). Because the leading logarithm approximation seems to 
be a well-defined concept for any Green's function we might suspect that 
it can be implemented at the level of the Heisenberg field equations.

Starobinski\u{\i} has long argued that his stochastic field equations
\cite{AAS2} should recover the leading infrared logarithms of any Green's
function at each order in perturbation theory. In 1994 he and Yokoyama 
\cite{SY} were able to obtain an explicit late time solution of these
stochastic equations for (\ref{Lphi}) which agrees with the intuitive
expectation that the growth of its field strength is eventually halted 
by the classical restoring force of the potential. The purpose of this 
paper is to generalize Starobinski\u{\i}'s technique so that it can be 
applied to models with derivative interactions and constrained fields 
such as gravity.

\section{The Physics of Infrared Logarithms}

The origin of the infrared logarithms in expressions (\ref{rhoGR}-\ref{pGR})
and (\ref{rhophi}-\ref{pphi}) can be understood physically as well as 
from the mathematics of perturbation theory. Although the physical
understanding is vastly more important in guiding the generalization 
we must make, it is crucial to see that the formalism provides an invariant 
separation between infrared and ultraviolet degrees of freedom. We
accordingly begin by explaining how perturbative computations are done
invariantly in a locally de Sitter background.

We employ dimensional regularization in position space. Of course the
vertices are straightforward to obtain in any background, so the only
difficulty comes in finding the propagators in arbitrary spacetime 
dimension $D$. These propagators are expressed in terms of a de Sitter 
invariant length function we call $y(x;x')$,
\begin{equation}
y(x;x') \equiv a a' \Biggl[ H^2 \Bigl\Vert \vec{x} - \vec{x}' 
\Bigr\Vert^2 - \Bigl(-\frac1{a} + \frac1{a'}\Bigr)^2\Biggr] \; .
\end{equation}
Its physical meaning in terms of the invariant length $\ell(x;x')$ 
between $x^{\mu}$ and $x^{\prime \mu}$ is,
\begin{equation}
y(x;x') = 4 \sin^2\Biggl[\frac12 H \ell(x;x') \Biggr] \; .
\end{equation}
For example, the propagator of a massless, conformally coupled scalar 
is \cite{BD},
\begin{equation}
{i\Delta}_{\rm cf}(x;x') = \frac{H^{D-2}}{(4\pi)^{\frac{D}2}} \Gamma\Bigl(
\frac{D}2 \!-\! 1\Bigr) \Bigl(\frac4{y}\Bigr)^{\frac{D}2-1} \; . \label{Dcon}
\end{equation}
Because $y(x;x')$ vanishes when $x^{\prime \mu} = x^{\mu}$, and because
one always interprets $D$ so that zero is raised to a positive power
in dimensional regularization, the coincidence limit of the conformally
coupled propagator vanishes,
\begin{equation}
{i\Delta}_{\rm cf}(x;x) = 0 \; .
\end{equation}

Of course conformal invariance suppresses interesting quantum effects
but $i\Delta_{\rm cf}(x;x')$ is useful for expressing the propagator of
the minimally coupled scalar \cite{OW1,OW2},
\begin{eqnarray}
\lefteqn{i \Delta_A(x;x') =  i \Delta_{\rm cf}(x;x') 
+ \frac{H^{D-2}}{(4\pi)^{\frac{D}2}} \frac{\Gamma(D \!-\! 1)}{\Gamma(
\frac{D}2)} \left\{\! \frac{D}{D\!-\! 4} \right. } \nonumber \\
& & \left. \times \frac{\Gamma^2(\frac{D}2)}{\Gamma(D
\!-\! 1)} \Bigl(\frac4{y}\Bigr)^{\frac{D}2 -2} \!\!\!\!\!\! - \pi 
\cot\Bigl(\frac{\pi}2 D\Bigr) + \ln(a a') \!\right\} \nonumber \\
& & + \frac{H^{D-2}}{(4\pi)^{\frac{D}2}} \! \sum_{n=1}^{\infty}\! \left\{\!
\frac1{n} \frac{\Gamma(n \!+\! D \!-\! 1)}{\Gamma(n \!+\! \frac{D}2)} 
\Bigl(\frac{y}4 \Bigr)^n \right. \nonumber \\
& & \left. - \frac1{n \!-\! \frac{D}2 \!+\! 2} \frac{\Gamma(n \!+\!  \frac{D}2 
\!+\! 1)}{\Gamma(n \!+\! 2)} \Bigl(\frac{y}4 \Bigr)^{n - \frac{D}2 +2} 
\!\right\} \! . \label{DeltaA}
\end{eqnarray}
This expression might seem daunting but it is actually simple to use in 
low order computations because the infinite sum on the final line vanishes 
in $D=4$, and each term in the series goes like a positive power of $y(x;x')$. 
This means that the infinite sum can only contribute when multiplied by a 
divergence, and even then only a small number of terms can contribute.

The explicit factor of $\ln(a a')$ at the end of the second line in
(\ref{DeltaA}) is one source of infrared logarithms. It gives the secular 
dependence of the coincidence limit \cite{VF,L,AAS3},
\begin{eqnarray}
\lefteqn{{i\Delta}_A(x;x) = \frac{H^{D-2}}{(4\pi)^{\frac{D}2}} } \nonumber \\
& & \times \frac{\Gamma(D \!-\! 1)}{\Gamma( \frac{D}2)} \left\{\! - \pi 
\cot\Bigl(\frac{\pi}2 D\Bigr) + 2 \ln(a) \!\right\} . \label{coinc}
\end{eqnarray}
This factor of $\ln(a a')$ is also interesting in that it breaks de Sitter
invariance. Of course $\varphi(x)$ transforms like a scalar; the breaking
derives from the state in which the expectation value of the two free fields
is taken. Allen and Folacci long ago proved that the massless, minimally 
coupled scalar fails to possess normalizable de Sitter invariant states 
\cite{AF}. A final point is that the factor of $\ln(a a')$ is actually an
invariant, it just depends upon the initial value surface at which the
state is released. In fact it is the Hubble constant times the sum of the 
invariant times from $x^{\mu}$ and $x^{\prime \mu}$ to this initial value 
surface. The physical reason for such a term is the increasing field
amplitude due to inflationary particle production, as discussed in the
introduction.

The graviton propagator is much more complicated on account of its
tensor indices. However, it also involves the $A$-type propagator 
\cite{TW4,RPW} so it can also induce infrared logarithms. Any infrared 
logarithms which appear in one loop diagrams involving scalars and/or
gravitons can only have come from this $A$-type propagator. However, 
another mechanism can produce infrared logarithms in higher loop results 
such as (\ref{rhoGR}-\ref{pGR}) and (\ref{rhophi}-\ref{pphi}). This 
other mechanism is the growth of the invariant volume of the past 
light-cone from the observation point back to the initial value 
surface,
\begin{eqnarray}
\lefteqn{V_{\mbox{\tiny PLC}}(t) \equiv \int_0^t dt' 
a^{\prime {\scriptscriptstyle D-1}} \int d^{\scriptscriptstyle D-1}x' 
\, \theta\Bigl[-y(x;x')\Bigr] } \nonumber \\
& & = \frac{2 \, \pi^{\frac{D-1}2} H^{-D}}{(D \!-\! 1) \Gamma(\frac{D-1}2)} 
\Bigl[\ln(a) + O(1)\Bigr] . \label{Vplclimit}
\end{eqnarray}
Factors of this quantity arise naturally whenever undifferentiated 
propagators connect an interaction vertex (at $x^{\prime \mu}$)
with the expectation value of some observable (at $x^{\mu}$). 

It should be noted that one must use the Schwinger-Keldysh formalism 
\cite{JS} to obtain true expectation values for cases, such as 
cosmology, in which the ``in'' and ``out'' vacua either do not agree 
or are not even well defined \cite{RJ}. In the Schwinger-Keldysh
formalism the only net contributions arise from interaction vertices 
which lie on or within the past light-cone of some observation point.
The more familiar in-out matrix elements would harbor virulent 
infrared divergences from integrating over the exponentially large 
volume of the inflationary future \cite{TW5}. The causality of the 
Schwinger-Keldysh formalism regulates these infrared divergences 
but simple correspondence implies that the regulated expressions must 
grow without bound at late times.

To summarize, infrared logarithms de\-rive joint\-ly from explicit factors 
of $\ln(a a')$ in the propagators and from integrating undifferentiated 
propagators over the past light-cone. Our reason for discussing the
mathematics of perturbation theory is to emphasize that the ultraviolet
regularization is invariant. This has a crucially important consequence
for understanding the physical origin of infrared logarithms: it means
that there should be a constant dynamical impact from all modes whose
physical wavelength ranges from zero (the far ultraviolet) to any fixed 
value. 

Recall from section 2 that quanta are labeled by constant wave 
vectors $\vec{k}$, and that the mode with wavenumber $k = \Vert \vec{k} 
\Vert$ begins to experience significant inflationary particle production 
when $k/a(t) = H$. This suggests that we distinguish ``infrared'' and 
``ultraviolet'' modes on the following basis,
\begin{eqnarray}
{\it Infrared} \quad & \Longrightarrow & \quad H < k < H \, a(t)
\;\; , \label{irmodes} \\
{\it Ultraviolet} 
\quad & \Longrightarrow & \quad k > H \, a(t) \;\; . \label{uvmodes}
\end{eqnarray}
The ultraviolet modes are those whose physical wavelengths 
($= 2\pi a(t)/k$) range from zero to the constant $2\pi/H$. With any
invariant regularization the dynamical impact of such modes must be
constant because they lie in an invariantly defined range. 

Had we taken the lower limit in expression (\ref{irmodes}) to $k=0$
the infrared phase space would also have extended over a constant 
physical range. However, taking $k$ down to zero has long been known 
to result in infrared divergences \cite{FP}. We regulate these by
working on $T^{D-1} \times R$, with the coordinate toroidal radii 
equal to $2\pi/H$ \cite{TW5}. (We also typically make the integral 
approximation to discrete mode sums.) There are other techniques 
\cite{FSW} but they all cause the effective infrared phase space to
increase as the universe inflates. This growth is the physical source 
of infrared logarithms. Essentially what happens is that the average
field strength increases as it receives contributions from more and 
more infrared modes. If any interactions involve the undifferentiated 
field then this growth can enter physical quantities.

These assertions are easy to confirm in the context of the scalar 
$\varphi^4$ model (\ref{Lphi}) in $D = 3 + 1$ dimensions. The field 
equation is,
\begin{equation}
\ddot{\varphi} + 3 H \dot{\varphi} - \frac{\nabla^2}{a^2} \varphi
+ \frac{\lambda}{6} \varphi^3 = 0 \; . \label{fulleqn}
\end{equation}
The full (ultraviolet plus infrared) perturbative initial value solution 
is obtained by iterating the following equation,
\begin{eqnarray}
\lefteqn{\varphi(t,\vec{x}) = \varphi_0(t,\vec{x}) - \frac{\lambda}6
\int_0^t dt' a^3(t') } \nonumber \\
& & \hspace{.5cm} \times \int d^3x' G_{\rm ret}\Bigl(t,\vec{x};t',\vec{x}'\Bigr)
\varphi^3(t',\vec{x}') . 
\end{eqnarray}
The retarded Green's function is,
\begin{eqnarray}
\lefteqn{G_{\rm ret}\Bigl(t,\vec{x};t',\vec{x}'\Bigr) } \nonumber \\
& & \hspace{-.5cm} \equiv \frac{H^2}{4\pi} \theta(t\!-\!t') \Biggl\{ 
\frac{\delta\Bigl(H \Vert\vec{x} \!-\! \vec{x}' \Vert + \frac1{a} - 
\frac1{a'}\Bigr)}{a a' H \Vert \vec{x} \!-\! \vec{x}' \Vert} \nonumber \\
& & \hspace{1.5cm} + \theta\Bigl(H \Vert\vec{x} \!-\! \vec{x}' \Vert + 
\frac1{a} - \frac1{a'}\Bigr) \Biggr\}\; .
\end{eqnarray}
The free field is,
\begin{equation}
\varphi_0(t,\vec{x}) = \int \!\! \frac{d^3k}{(2\pi)^3} \Bigl\{ 
e^{i \vec{k} \cdot \vec{x}} u(t,k) \alpha(\vec{k}) + {\rm c.c.} \Bigr\} , 
\label{free}
\end{equation}
where the mode function and commutation relations are,
\begin{eqnarray}
& & u(t,k) = \frac{H}{\sqrt{2 k^3}} \Bigl[1 - \frac{i k}{H a}\Bigr] 
e^{\frac{ik}{Ha}} \nonumber \\
& & {\rm and} \quad \Bigl[\alpha(\vec{k}),\alpha^{\dagger}(\vec{k}')\Bigr] 
= (2\pi)^3 \delta^3\Bigl(\vec{k} \!-\! \vec{k}'\Bigr) . \label{mcrs}
\end{eqnarray}
Note that although $\varphi_0(t,\vec{x})$ is only the order $\lambda^0$
part of the solution, it and its first derivative agree exactly with 
the full field on the initial value surface.

To excise the ultraviolet modes (\ref{uvmodes}) we iterate what is 
essentially the same equation,
\begin{eqnarray}
\lefteqn{\Phi(t,\vec{x}) = \Phi_0(t,\vec{x}) - \frac{\lambda}6 \int_0^t 
dt' a^3(t') } \nonumber \\
& & \hspace{.5cm} \times \int d^3x' G_{\rm ret}\Bigl(t,\vec{x};t',\vec{x}'\Bigr)
\Phi^3(t',\vec{x}') \; , \label{fullIR}
\end{eqnarray}
but with the zeroth order solution restricted to only infrared modes,
\begin{eqnarray}
\lefteqn{\Phi_0(t,\vec{x}) = \int \!\! \frac{d^3k}{(2\pi)^3} \, 
\theta\Bigl(Ha(t) \!-\! k\Bigr) } \nonumber \\
& & \hspace{-.5cm} \times \Bigl\{ e^{i \vec{k} \cdot \vec{x}} u(t,k) 
\alpha(\vec{k}) \!+\! e^{-i \vec{k} \cdot \vec{x}} u^*(t,k) 
\alpha^{\dagger}(\vec{k}) \Bigr\} , 
\end{eqnarray}
This model is completely free of ultraviolet divergences, so we are 
justified in taking $D = 3 + 1$. To see that the model also reproduces 
the leading infrared logarithms at tree order it suffices to take the 
expectation value of $\Phi_0^2$,
\begin{eqnarray}
\lefteqn{\Bigl\langle \Omega \Bigl\vert \Phi^2_0(t,\vec{x}) \Bigr\vert \Omega 
\Bigr\rangle } \nonumber \\
& & = \int \! \frac{d^3k}{(2\pi)^3} \, \theta\Bigl(H a
- k\Bigr) \frac{H^2}{2 k^3} \Bigg\{1 + \frac{k^2}{H^2 a^2} \Biggr\} , \\
& & = \Biggl(\frac{H}{2\pi} \Biggr)^2 \Biggl\{\ln(a) + \frac12 - 
\frac1{2 a^2} \Biggr\} . \label{P0sq}
\end{eqnarray}
Comparison with (\ref{coinc}) (for $D= 3+1$) reveals exact agreement
between the $\ln(a)$ terms.

Our purely infrared field operator $\Phi(t,\vec{x})$ does not quite obey 
the field equation (\ref{fulleqn}) because the kinetic operator fails to
annihilate $\Phi_0(t,\vec{x})$. When time derivatives act upon the time 
dependent upper limit of the mode sum,
\begin{eqnarray}
\lefteqn{\ddot{\Phi}_0(t,\vec{x}) + 3 H \dot{\Phi}_0(t,\vec{x}) - 
\frac{\nabla^2}{a^2(t)} \Phi_0(t,\vec{s}) } \nonumber \\
& & = \dot{\mathcal{F}}(t,\vec{x}) + \mathcal{G}(t,\vec{x}) + 3 H 
\mathcal{F}(t,\vec{x}) , \label{extra}
\end{eqnarray}
it produces momentum space surface terms involving the sources,
\begin{eqnarray}
\lefteqn{\mathcal{F}(t,\vec{x}) = H \!\!\int \!\! \frac{d^3k}{(2\pi)^3} \, 
k \delta\Bigl(k \!-\! H a(t)\Bigr) } \nonumber \\
& & \hspace{-.5cm} \times \Bigl\{ e^{i \vec{k} \cdot \vec{x}} u(t,k) 
\alpha(\vec{k}) \!+\! e^{-i \vec{k} \cdot \vec{x}} u^*(t,k) 
\alpha^{\dagger}(\vec{k}) \Bigr\} , \\
\lefteqn{\mathcal{G}(t,\vec{x}) = H \!\!\int \!\! \frac{d^3k}{(2\pi)^3} \, 
k \delta\Bigl(k \!-\! H a(t)\Bigr) } \nonumber \\
& & \hspace{-.5cm} \times \Bigl\{ e^{i \vec{k} \cdot \vec{x}} \dot{u}(t,k) 
\alpha(\vec{k}) \!+\! e^{-i \vec{k} \cdot \vec{x}} \dot{u}^*(t,k) 
\alpha^{\dagger}(\vec{k}) \Bigr\} .
\end{eqnarray}
Note that the mode functions and their derivatives are simply constants
at $k = H a(t)$,
\begin{eqnarray}
& & u(t,k) \Bigl\vert_{k = H a(t)} = \frac{H}{\sqrt{2 k^3}} \Bigl(1 - i\Bigr)
e^i , \\
& & \dot{u}(t,k) \Bigl\vert_{k = H a(t)} = -\frac{H^2}{\sqrt{2 k^3}} e^i 
\end{eqnarray}

In view of (\ref{extra}) the equation obeyed by $\Phi(t,\vec{x})$ is not
(\ref{fulleqn}) but rather,
\begin{equation}
\Bigl(\ddot{\Phi} - \dot{\mathcal{F}} - \mathcal{G} \Bigr) + 3 H \Bigl(
\dot{\Phi} - \mathcal{F}\Bigr) - \frac{\nabla^2}{a^2} \Phi +
\frac{\lambda}{6} \Phi^3 = 0 \; . \label{Phieqn}
\end{equation}
This equation does not leave the original stress-energy tensor conserved
because stress-energy is being continually dumped into the truncated
system by ultraviolet modes which redshift past the horizon. We can
account for this by modifying what we call $T_{\mu\nu}$. To motivate the
modification it is useful to write down the form we expect for the 
divergence of the stress-energy. For the zero component we should get,
\begin{eqnarray}
\lefteqn{T_{0 \mu}^{~~;\mu} = -\dot{T}_{00} - H \Bigl[3 T_{00} + g^{ij} T_{ij}
\Bigr] + g^{ij} T_{0i,j} ,} \\
& & = -\dot{\Phi} \Biggl[ \Bigl(\ddot{\Phi} - \dot{\mathcal{F}} -
\mathcal{G}\Bigr) + 3 H \Bigl(\dot{\Phi} - \mathcal{F}\Bigr) \nonumber \\
& & \hspace{3.5cm} - \frac{\nabla^2}{a^2} \Phi + \frac{\lambda}{6} \Phi^3 
\Biggr] . \label{timeT}
\end{eqnarray}
The spatial components should read,
\begin{eqnarray}
\lefteqn{T_{i \mu}^{~~;\mu} = -\dot{T}_{0i} - 3 H T_{0i} + \frac1{a^2} T_{ij,j}
\; , } \\
& & = -\partial_i \Phi \Biggl[ \Bigl(\ddot{\Phi} - \dot{\mathcal{F}}
- \mathcal{G}\Bigr) + 3 H \Bigl(\dot{\Phi} - \mathcal{F}\Bigr) \nonumber \\
& & \hspace{3.5cm} - \frac{\nabla^2}{a^2} \Phi + \frac{\lambda}{6} \Phi^3 
\Biggr] . \label{spaceT}
\end{eqnarray}

We can enforce (\ref{timeT}) and (\ref{spaceT}) with a stress-energy of the
form,
\begin{eqnarray}
\lefteqn{T_{00} = \frac12 \Bigl(\dot{\Phi} \!-\! \mathcal{F}\Bigr)^2 + 
\frac1{2a^2} \vec{\nabla} \Phi \!\cdot\! \vec{\nabla} \Phi + 
\frac{\lambda}{24} \Phi^4 , } \\
\lefteqn{T_{0i} = \Bigl(\dot{\Phi} \!-\! \mathcal{F}\Bigr) \partial_i 
\Phi , } \\
\lefteqn{T_{ij} = \partial_i \Phi \partial_j \Phi - g_{ij} \Biggl[-\frac12 
\Bigl(\dot{\Phi} - \mathcal{F}\Bigr)^2 + \frac1{2 a^2} \vec{\nabla} 
\Phi \!\cdot\! \vec{\nabla} \Phi} \nonumber \\
& & + \frac{\lambda}{24} \Phi^4\Biggr] + \partial_i S_j + \partial_j S_i 
- \frac12 \Bigl(\delta_{ij} \!-\! 3 \frac{\partial_i \partial_j}{\nabla^2}
\Bigr) S^L \nonumber \\
& & \hspace{3cm} + \frac12 \Bigl(\delta_{ij} \!-\! \frac{\partial_i 
\partial_j}{\nabla^2}\Bigr) S \; .
\end{eqnarray}
Once this form is assumed the nonlocal source terms of the purely spatial 
components are determined by conservation to be,
\begin{eqnarray}
\lefteqn{S_i = \frac{a^2}{\nabla^2} \Bigl[\mathcal{G} \partial_i \Phi + 
(\dot{\Phi} \!-\! \mathcal{F}) \partial_i \mathcal{F} \Bigr] } \nonumber \\
& & - \frac{a^4 \partial_i \partial_k}{\nabla^4} \Bigl[\mathcal{G} \partial_k
\Phi + (\dot{\Phi} \!-\! \mathcal{F}) \partial_k \mathcal{F} \Bigr], \\
\lefteqn{S^L = \frac{a^2 \partial_k}{\nabla^2} \Bigl[ \mathcal{G}
\partial_k \Phi + (\dot{\Phi} \!-\! \mathcal{F}) \partial_k \mathcal{F} 
\Bigr] , } \\
\lefteqn{S = -\frac{a^2}{H} \Bigl(\dot{\Phi} \!-\! \mathcal{F}\Bigr) 
\mathcal{G} -\frac1{H} \vec{\nabla} \mathcal{F} \cdot \vec{\nabla} \Phi -
\frac{\lambda a^2}{6 H} \mathcal{F} \Phi^3 .}
\end{eqnarray}
We have therefore achieved a completely consistent model of just the
infrared modes that reproduces the leading infrared logarithms.

Although the zeroth order field $\Phi_0(t,\vec{x})$ contains only
infrared modes it is still quantum mechanical in that the field and
its first time derivative do not commute,
\begin{eqnarray}
\lefteqn{\Bigl[\Phi_0(t,\vec{x}),\dot{\Phi}_0(t,\vec{x}')\Bigr] = \int 
\frac{d^3k}{(2 \pi)^3} \, \theta\Bigl(H a(t) \!-\! k\Bigr) } \nonumber \\
& & \times \Bigl[u(t,k) \dot{u}^*(t,k) - u^*(t,k) \dot{u}(t,k)\Bigr] 
e^{i \vec{k} \cdot \Delta \vec{x}} , \\
& & = \frac{i}{4 \pi^2 a^3} \int_H^{Ha} \! dk \, k^2 \frac{\sin(k 
\Delta x)}{k \Delta x} , \\
& & = \frac{i}{4 \pi^2 (a \Delta x)^3} \nonumber \\
& & \times \left\{ \matrix{\sin(a H \Delta x) - a H \Delta x \cos(a H 
\Delta x) \cr - \sin(H \Delta x) + H \Delta x \cos(H \Delta x)} \right\} .
\end{eqnarray}
(In these formulae we define $\Delta \vec{x} \equiv \vec{x} - \vec{x}'$
and $\Delta x \equiv \Vert \Delta \vec{x} \Vert$.) However, note that 
the leading infrared logarithm in (\ref{P0sq}) derives entirely from 
the constant first term of the long wavelength expansion of the mode 
function,
\begin{equation}
u(t,k) \!=\! \frac{H}{\sqrt{2k^3}} \Biggl\{\!1 \!+\! \frac12 
\Bigl(\!\frac{k}{Ha}\!\Bigl)^{\!2} \!+ \frac{i}3
\Bigl(\!\frac{k}{Ha}\!\Bigl)^{\!3} \!+\! \dots \!\Biggr\} . \label{long}
\end{equation}
We would get precisely the same infrared logarithm from a free field with 
only this term,
\begin{eqnarray}
\lefteqn{\phi_0(t,\vec{x}) \equiv \int \!\! \frac{d^3k}{(2\pi)^3} \, 
\theta\Bigl(Ha(t) \!-\! k\Bigr) } \nonumber \\
& & \times \frac{H}{\sqrt{2 k^3}} \Bigl\{ e^{i \vec{k} \cdot \vec{x}} 
\alpha(\vec{k}) + e^{-i \vec{k} \cdot \vec{x}} \alpha^{\dagger}(\vec{k}) 
\Bigr\} . \label{freephi}
\end{eqnarray}
However, the field $\phi_0(t,\vec{x})$ is ``classical'' in the sense that
it commutes with its time derivative. Because it is a superposition of 
creation and annihilation operators $\phi_0(t,\vec{x})$ is still a quantum 
operator and hence liable to take any value. Such a random but commuting 
variable might be termed ``stochastic.''

How should one include corrections to $\phi_0(t,\vec{x})$? We could
simply replace $\Phi_0$ by $\phi_0$ in (\ref{fullIR}). Iterating such 
a relation would capture the leading infrared logarithms however, we 
may as well simplify the Green's function along the same lines as the
mode function. The full Green's function can be expressed in terms of
the full mode function as follows,
\begin{eqnarray}
\lefteqn{G_{\rm ret}\Bigl(t,\vec{x};t',\vec{x}'\Bigr) = i \theta(t-t') \int
\frac{d^3k}{(2\pi)^3} } \nonumber \\
& & \hspace{-.5cm} \times \Bigl[u(t,k) u^*(t',k) - u^*(t,k) u(t',k)\Bigr] 
e^{i \vec{k} \cdot \Delta \vec{x}} . \label{Gmodes}
\end{eqnarray}
The long wavelength expansion of the bracketed term is,
\begin{eqnarray}
\lefteqn{\Bigl[u(t,k) u^*(t',k) - u^*(t,k) u(t',k)\Bigr] } \nonumber \\
& & = \frac{i}{3H} \Biggl(\frac1{a^3} - \frac1{a^{\prime 3}} \Biggr) 
+ O(k^2) .
\end{eqnarray}
Retaining only this leading term in (\ref{Gmodes}) gives,
\begin{eqnarray}
\lefteqn{G_{\rm ret}\Bigl(t,\vec{x};t',\vec{x}'\Bigr) } \nonumber \\
& & \longrightarrow \frac1{3H} \theta(t-t') \delta^3(\vec{x} - \vec{x}') 
\Biggl( \frac1{a^{\prime 3}} - \frac1{a^3} \Biggr) . \label{Glong}
\end{eqnarray}
It is actually superfluous even keeping the $1/a^3$ term, which can 
never result in a leading infrared logarithm. We will therefore recover
the leading infrared logarithms by iterating the simplified equation,
\begin{equation}
\phi(t,\vec{x}) = \phi_0(t,\vec{x}) - \frac{\lambda}{18 H} \int_0^t dt' 
\phi^3(t',\vec{x}) \; . \label{partIR}
\end{equation}

It remains to infer the local differential equation which the full
stochastic field $\phi(t,\vec{x})$ obeys. Note that the time derivative
of the stochastic free field is a momentum space surface term,
\begin{eqnarray}
\lefteqn{\dot{\phi}_0(t,\vec{x}) \equiv f(t,\vec{x}) , } \\
& & \hspace{-.5cm} = \!\! \int \!\! \frac{d^3k}{(2\pi)^3} \, \delta\Bigl(k 
\!-\! H a\Bigr) \frac{H^2}{\sqrt{2k}} \Bigl\{\!e^{i \vec{k} \cdot \vec{x}} 
\alpha(\vec{k}) \!+\! {\rm c.c.} \!\! \Bigr\} , 
\end{eqnarray}
Hence the full stochastic field obeys,
\begin{equation}
3 H \Bigl(\dot{\phi}(t,\vec{x}) - f(t,\vec{x})\Bigr) + \frac{\lambda}6
\phi^3(t,\vec{x}) = 0 \; . \label{stocheqn}
\end{equation}

\section{Derivative Interactions}

Of course (\ref{stocheqn}) is a Langevin equation with the white noise
provided by $f(t,\vec{x})$. For the techniques by which the associated 
Fokker-Planck equation for the probability density can be solved at 
asymptotically late times we recommend the excellent paper by 
Starobinski\u{\i} and Yokoyama \cite{SY}. Our main purpose is to 
derive the analogous Langevin equation for quantum gravity. This 
entails generalizing the derivation to cover derivative interactions 
(this section) and constrained fields (section 6).

The key property of the Langevin equation we seek is that it should
capture the leading infrared logarithms at each order in perturbation
theory. Because we shall have to make guesses about the equation it is 
desirable to be able to check these guesses against explicit perturbative 
computations in the full theory. The time required to do this in quantum 
gravity is prohibitive. We therefore studied a simple scalar model with 
derivative interactions,
\begin{eqnarray}
\lefteqn{\mathcal{L} = -\frac12 \partial_{\mu} A \partial_{\nu} A
g^{\mu\nu} \sqrt{-g} -\frac12 \partial_{\mu} B \partial_{\nu} B
g^{\mu\nu} \sqrt{-g} } \nonumber \\
& & \hspace{2cm} - \frac{\lambda}4 A^2 \partial_{\mu} B \partial_{\nu} B 
g^{\mu\nu} \sqrt{-g} . \label{Lder}
\end{eqnarray}
Both fields $A(t,\vec{x})$ and $B(t,\vec{x})$ are massless, minimally
coupled scalars at zeroth order, so both have the $A$-type propagator
(\ref{DeltaA}). With a little work we can evaluate the one loop
corrections to the square of each field \cite{TW6},
\begin{eqnarray}
\lefteqn{\Bigl\langle \Omega \Bigl\vert A^2(t,{\vec x}) \Bigr\vert \Omega 
\Bigr\rangle_{{\rm leading} \atop {\rm log}} = \frac{H^2}{4 \pi^2} 
\ln(a) + O(\lambda^2) ,} \label{derA2} \\
\lefteqn{\Bigl\langle \Omega \Bigl\vert B^2(t,{\vec x}) \Bigr\vert \Omega 
\Bigr\rangle_{{\rm leading} \atop {\rm log}} = \frac{H^2}{4 \pi^2} \ln(a) }
\nonumber \\
& & \hspace{2cm} - \frac{\lambda H^4}{2^6 \pi^4} \ln^2(a) + O(\lambda^2) . 
\label{derB2}
\end{eqnarray}

The full field equations for this system are,
\begin{eqnarray}
\lefteqn{\frac1{\sqrt{-g}} \frac{\delta S}{\delta A} = A^{;\mu}_{~~\mu} \!-\! 
\frac{\lambda}2 A B_{,\mu} B^{,\mu} = 0 ,} \label{derAeom} \\
\lefteqn{\frac1{\sqrt{-g}} \frac{\delta S}{\delta B} = B^{;\mu}_{~~\mu} \!+\! 
\lambda A A_{,\mu} B^{,\mu} \!+\! \frac{\lambda}2 A^2 B^{;\mu}_{~~\mu} 
= 0 . } \label{derBeom}
\end{eqnarray}
We can solve them perturbatively by iterating the following system,
\begin{eqnarray}
\lefteqn{A(t,\vec{x}) = A_{\rm fr}(t,\vec{x}) - \frac{\lambda}2 \! \int_0^t 
\! dt' a^{\prime 3} \! \int \! d^3x' G_{\rm ret}(x;x')} \nonumber \\
& & \hspace{2cm} \times A(x') B_{,\mu}(x') B^{,\mu}(x') , \\
\lefteqn{B(t,\vec{x}) = B_{\rm fr}(t,\vec{x}) + \lambda \! \int_0^t \! dt' 
a^{\prime 3} \! \int \! d^3x' G_{\rm ret}(x;x')} \nonumber \\
& & \hspace{-.6cm} \times \Bigl\{A(x') A_{,\mu}(x') B^{,\mu}(x') \!+\! 
\frac12 A^2(x') B^{;\mu}_{~~\mu}(x') \Bigr\} \! . 
\end{eqnarray}
(In these formulae we have sometimes compressed space and time arguments 
to a single 4-vector: $(t,\vec{x}) \equiv x$.) The two free fields are
independent copies of the scalar free field (\ref{free}),
\begin{eqnarray}
\lefteqn{A_{\rm fr}(t,\vec{x}) = \int \!\! \frac{d^3k}{(2\pi)^3} \Bigl\{ 
e^{i \vec{k} \cdot \vec{x}} u(t,k) \alpha(\vec{k}) + {\rm c.c.} \Bigr\} , } 
\label{Afr} \\
\lefteqn{B_{\rm fr}(t,\vec{x}) = \int \!\! \frac{d^3k}{(2\pi)^3} \Bigl\{ 
e^{i \vec{k} \cdot \vec{x}} u(t,k) \beta(\vec{k}) + {\rm c.c.} \Bigr\} . }
\end{eqnarray}
Here $\alpha(\vec{k})$ and $\beta(\vec{k})$ are independent, canonically
normalized annihilation operators and $u(t,k)$ is the mode function
(\ref{mcrs}) of the massless, minimally coupled scalar.

What is the simplified stochastic analogue that recovers the leading
infrared logarithms? From the analysis of the previous section it is
apparent that differentiating a propagator precludes it from contributing
an infrared logarithm. Said another way, it is always preferable to have 
a derivative act on the rapidly varying scale factor rather than the 
slowly varying field. We therefore expect that the Hubble friction term
dominates the scalar d'Alembertian,
\begin{equation}
A^{;\mu}_{~~\mu} = -\ddot{A} - 3 H \dot{A} + \frac{\nabla^2}{a^2} A
\longrightarrow -3 H \dot{A} .
\end{equation}
No term is comparably important when the two derivatives act on
different fields,
\begin{eqnarray}
A_{,\mu} B^{,\mu} &=& - \dot{A} \dot{B} + \frac1{a^2} \vec{\nabla} A \cdot 
\vec{\nabla} B \longrightarrow 0 , \\
B_{,\mu} B^{,\mu} &=& - \dot{B}^2 + \frac1{a^2} \vec{\nabla} B \cdot 
\vec{\nabla} B \longrightarrow 0 .
\end{eqnarray}

The rest of the derivation is identical to that of the previous section.
We again excise sub-horizon modes and take the long wavelength limit of
the mode function,
\begin{eqnarray}
\lefteqn{A_0(t,\vec{x}) \equiv \int \!\! \frac{d^3k}{(2\pi)^3} \, 
\theta\Bigl(Ha(t) \!-\! k\Bigr) } \nonumber \\
& & \times \frac{H}{\sqrt{2 k^3}} \Bigl\{ e^{i \vec{k} \cdot \vec{x}} 
\alpha(\vec{k}) + e^{-i \vec{k} \cdot \vec{x}} \alpha^{\dagger}(\vec{k}) 
\Bigr\} , \\
\lefteqn{B_0(t,\vec{x}) \equiv \int \!\! \frac{d^3k}{(2\pi)^3} \, 
\theta\Bigl(Ha(t) \!-\! k\Bigr) } \nonumber \\
& & \times \frac{H}{\sqrt{2 k^3}} \Bigl\{ e^{i \vec{k} \cdot \vec{x}} 
\beta(\vec{k}) + e^{-i \vec{k} \cdot \vec{x}} \beta^{\dagger}(\vec{k}) 
\Bigr\} .
\end{eqnarray}
We also take the long wavelength limit (\ref{Glong}) of the Green's 
function. Putting everything together gives the following simplified
iterative equations for reproducing the leading infrared logarithms,
\begin{eqnarray}
\lefteqn{A_{\mbox{\tiny IR}}(t,\vec{x}) = A_0(t,\vec{x}) , } \label{Asoln} \\
\lefteqn{B_{\mbox{\tiny IR}}(t,\vec{x}) = B_0(t,\vec{x}) } \nonumber \\
& & - \frac{\lambda}2 \int_0^t dt' A^2_{\mbox{\tiny IR}}(t',\vec{x})
\dot{B}_{\mbox{\tiny IR}}(t',\vec{x}) .
\end{eqnarray}

We differentiate to obtain local equations,
\begin{eqnarray}
-3 H \Bigl( \dot{A}_{\mbox{\tiny IR}} \!-\! f_{\mbox{\tiny A}} 
\Bigr) & = & 0 , \label{derAeqn} \\
-3 H \Bigl( \dot{B}_{\mbox{\tiny IR}} \!-\! f_{\mbox{\tiny B}} 
\Bigr) \!-\! \frac{3 \lambda H}{2} A_{\mbox{\tiny IR}}^2 
\dot{B}_{\mbox{\tiny IR}} & = & 0 . \label{derBeqn} 
\end{eqnarray}
Here the stochastic sources are,
\begin{eqnarray}
\lefteqn{f_{\mbox{\tiny A}}(t,\vec{x}) = \int \frac{d^3k}{(2\pi)^3} \, 
\delta\Bigl(k \!-\! H a(t)\Bigr) } \nonumber \\
& & \hspace{1cm} \times \frac{H^2}{\sqrt{2k}} \Bigl\{ e^{i \vec{k} 
\cdot \vec{x}} \alpha(\vec{k}) \!+\! e^{-i \vec{k} \cdot \vec{x}} 
\alpha^{\dagger}(\vec{k}) \Bigr\} , \label{fdefA} \\
\lefteqn{f_{\mbox{\tiny B}}(t,\vec{x}) = \int \frac{d^3k}{(2\pi)^3} \, 
\delta\Bigl(k \!-\! H a(t)\Bigr) } \nonumber \\
& & \hspace{1cm} \times \frac{H^2}{\sqrt{2k}} \Bigl\{ e^{i \vec{k} 
\cdot \vec{x}} \beta(\vec{k}) \!+\! e^{-i \vec{k} \cdot \vec{x}} 
\beta^{\dagger}(\vec{k}) \Bigr\} . \label{fdefB}
\end{eqnarray}
Note the curious fact that {\it even interactions which contain derivatives 
are free of stochastic source terms}. 

Although we would like to check it on other quantities and at higher 
orders, equations (\ref{derAeqn}-\ref{derBeqn}) do reproduce the known 
leading infrared logarithms (\ref{derA2}-\ref{derB2}). An amusing
feature of this system is that we can obtain an explicit operator 
solution rather than just a solution for the probability distribution
at asymptotically late times! Of course we can read off $A_{\mbox{\tiny IR}}$ 
from (\ref{Asoln}). A few simple rearrangements gives a closed form solution 
for $B_{\mbox{\tiny IR}}$ as well,
\begin{equation}
B_{\mbox{\tiny IR}}(t,\vec{x}) = \int_0^t dt' \,
\frac{f_{\mbox{\tiny B}}(t',\vec{x})}{1 \!+\! \frac12 \lambda 
A^2_0(t', \vec{x})} . \label{Bsoln}
\end{equation}
It exhibits a sort of spacetime-dependent field strength renormalization
whereby each contingent of $\beta$-modes which experiences horizon crossing
is attenuated by the factor $1 + \frac{\lambda}2 A_0^2(t,\vec{x})$.

\section{Constrained Fields}

Gravity also possesses constrained fields. We might anticipate that these
possess no stochastic source terms because they have no independent 
degrees of freedom. Finding a scalar analogue model in which to check
this is difficult because pure scalar models lack gauge constraints.
However, it is possible to go the other way by formulating gravity in
a covariant gauge \cite{AW} for which, in $D=4$ spacetime dimensions, all 
the free fields are either massless, minimally coupled scalars or else
massless, conformally coupled scalars. We can then compare scalar models
involving the two sorts of fields.

A simple model of the desired type is,
\begin{eqnarray}
\lefteqn{\mathcal{L} = -\frac12 \partial_{\mu} A \partial_{\nu} A
g^{\mu\nu} \sqrt{-g} \!-\! \frac12 \partial_{\mu} C \partial_{\nu} C
g^{\mu\nu} \sqrt{-g} } \nonumber \\
& & -\frac{D \!-\! 2}{8 (D \!-\! 1)} C^2 R \sqrt{-g} 
\!-\! \frac12 \kappa H^2 A^2 C \sqrt{-g} . \label{Lcon}
\end{eqnarray}
The $A$ field again has the $A$-type propagator (\ref{DeltaA}) while the
$C$ field has the conformal propagator (\ref{Dcon}). Taking the VEV's of 
the squares of each field gives,
\begin{eqnarray}
\lefteqn{\Bigl\langle \Omega \Bigl\vert A^2(t,{\vec x}) \Bigr\vert \Omega 
\Bigr\rangle_{{\rm leading} \atop {\rm log}} = \frac{H^2}{4 \pi^2} 
\ln(a) + O(\kappa^2) .} \label{conA2} \\
\lefteqn{\Bigl\langle \Omega \Bigl\vert C^2(t,{\vec x}) \Bigr\vert \Omega 
\Bigr\rangle_{{\rm leading} \atop {\rm log}} = 0 + O(\kappa^2) .} 
\label{conB2}
\end{eqnarray}

The exact field equations are,
\begin{eqnarray}
\lefteqn{\frac1{\sqrt{-g}} \, \frac{\delta S}{\delta A} = 
A^{;\mu}_{~~\mu} \!-\! \kappa H^2 A C = 0 ,} \label{LconAeom} \\
\lefteqn{\frac1{\sqrt{-g}} \, \frac{\delta S}{\delta C} =
C^{;\mu}_{~~\mu} \!-\! 2 H^2 C \!-\! \frac12 \kappa H^2 A^2 = 0 .}
\label{LconBeom}
\end{eqnarray}
As usual the perturbative solution follows by interating the
integrated field equations,
\begin{eqnarray}
\lefteqn{A(t,\vec{x}) = A_{\rm fr}(t,\vec{x}) - \kappa H^2 \int_0^t 
dt' a^3(t') \int d^3x' } \nonumber \\
& & \hspace{1cm} \times G_{\rm ret}\Bigl(t,\vec{x};t',\vec{x}'\Bigr)
A(t',\vec{x}') C(t',\vec{x}') , \\
\lefteqn{C(t,\vec{x}) = C_{\rm fr}(t,\vec{x}) - \frac12 \kappa H^2 \int_0^t 
dt' a^3(t') \int d^3x' } \nonumber \\
& & \hspace{1cm} \times \Gamma_{\rm ret}\Bigl(t,\vec{x};t',\vec{x}'\Bigr)
A^2(t',\vec{x}') .
\end{eqnarray}
The free $A$-type field is identical to (\ref{Afr}). However, the free
$C$-type field has a different mode function,
\begin{eqnarray}
\lefteqn{C_{\rm fr}(t,\vec{x}) = \int \!\! \frac{d^3k}{(2\pi)^3} \Bigl\{ 
e^{i \vec{k} \cdot \vec{x}} v(t,k) \gamma(\vec{k}) + {\rm c.c.} \Bigr\} .} \\
\lefteqn{v(t,k) \equiv \frac{H}{\sqrt{2 k^3}} \Bigl(\frac{-i}{H a}\Bigr)
e^{\frac{ik}{H a}} = \frac{-i}{\sqrt{2 k}} \, \frac1{a} \,
e^{\frac{ik}{H a}} . } \label{cmod}
\end{eqnarray}
The $C$-type field also has a different retarded Green's function,
\begin{eqnarray}
\lefteqn{\Gamma_{\rm ret}\Bigl(t,\vec{x};t',\vec{x}'\Bigr) } \nonumber \\
& & \equiv \frac{H^2}{4\pi} \, \theta(t\!-\!t') \,
\frac{\delta\Bigl(H \Vert\vec{x} \!-\! \vec{x}' \Vert + \frac1{a} - 
\frac1{a'}\Bigr)}{a a' H \Vert \vec{x} \!-\! \vec{x}' \Vert} .
\end{eqnarray}

We can implement the leading logarithm approximation on the $A$-type
field equation using techniques which are by now familiar,
\begin{eqnarray}
\lefteqn{A_{\mbox{\tiny IR}}(t,\vec{x}) = A_0(t,\vec{x}) } \nonumber \\
& & - \frac13 \kappa H \int_0^t dt' A_{\mbox{\tiny IR}}(t',\vec{x})
C_{\mbox{\tiny IR}}(t',\vec{x}) .
\end{eqnarray}
Achieving a similar reduction for the $C$ equation requires several
new insights. First, the $C$-type mode function (\ref{cmod}) lacks the 
$k^{-\frac32}$ singularity needed to produce infrared logarithms. It 
follows that the $C$-type free field truncates to zero in the leading 
logarithm approximation! The long wavelength limit of the $C$-type 
retarded Green's function is also different,
\begin{eqnarray}
\lefteqn{\Gamma_{\rm ret}\Bigl(t,\vec{x};t',\vec{x}'\Bigr) } \nonumber \\
& & \hspace{-.5cm} \longrightarrow \frac1{H} \theta(t-t') \delta^3(\vec{x} 
- \vec{x}') \Biggl(\frac1{a a^{\prime 2}} - \frac1{a^2 a'} \Biggr) . 
\end{eqnarray}
These insights allow us to write a leading logarithm approximation for
the $C$ equation as follows,
\begin{equation}
C_{\mbox{\tiny IR}}(t,\vec{x}) \!=\! -\frac{\kappa H}2 \!\! \int_0^t \!\! dt' 
\Bigl(\frac{a'}{a} - \frac{a^{\prime 2}}{a^2} \Bigr)
A^2{\mbox{\tiny IR}}(t',\vec{x}) .
\end{equation}
A final simplification is to integrate by parts on the rapidly varying
scale factors and neglect the slow variation in $A_{\mbox{\tiny IR}}(t',
\vec{x})$ to obtain a local equation,
\begin{equation}
C_{\mbox{\tiny IR}} = -\frac14 \kappa A_{\mbox{\tiny IR}}^2 . \label{Csoln} 
\end{equation}

We can rewrite the simplified leading logarithm field equations as follows,
\begin{eqnarray}
-3H \Bigl( \dot{A}_{\mbox{\tiny IR}} \!-\! f_{\mbox{\tiny A}} \Bigr)
- \kappa H^2 A_{\mbox{\tiny IR}} C_{\mbox{\tiny IR}} & = & 0 , 
\label{conAeqn} \\
-2H^2 C_{\mbox{\tiny IR}} - \frac12 \kappa H^2 A^2_{\mbox{\tiny IR}} 
& = & 0 . \label{conCeqn} 
\end{eqnarray}
The $C$ equation (\ref{conCeqn}) could be derived from (\ref{LconBeom}) 
by arguing that it is preferable to act derivatives on the rapidly varying 
scale factor rather than the slowly varying field. Hence the conformal 
d'Alembertian is dominated by the curvature term,
\begin{eqnarray}
\lefteqn{C^{;\mu}_{~~\mu} - \frac16 R C = -\ddot{C} - 3 H \dot{C} + 
\frac{\nabla^2}{a^2} C - 2 H^2 C } \nonumber \\
& & \longrightarrow -2 H^2 C .
\end{eqnarray}

We would again like to check more expectation values and higher orders,
but our putative equations (\ref{conAeqn}-\ref{conCeqn}) do reproduce
the known leading infrared logarithms (\ref{conA2}-\ref{conB2}).
Substituting (\ref{Csoln}) into the $A$ equation (\ref{conAeqn}) gives,
\begin{equation}
\dot{A}_{\mbox{\tiny IR}} = f_{\mbox{\tiny A}} - \frac{\kappa^2 H}{12} 
A_{\mbox{\tiny IR}}^3 . \label{conAeqn2} 
\end{equation}
It follows that $A_{\mbox{\tiny IR}}(t,\vec{x})$ and $\phi(t,\vec{x})$
obey the same equation with $\lambda = \frac32 \kappa^2 H^2$!

\section{Conclusions}

We can at length state what seem to be the general rules for
inferring simplified operator equations which reproduce the
leading infrared logarithms:
\begin{enumerate}
\item{At each order in the field (e.g., $\varphi^1$, $\varphi^2$,
etc.) retain only the term with the smallest number of derivatives.}
\item{Each time derivative in the linear terms has a stochastic
source subtracted.}
\end{enumerate}
Of course we would get the same leading infrared logarithms by
solving (\ref{Phieqn}) for $\Phi(t,\vec{x})$. The enormous advantage
of solving the stochastic equation (\ref{stocheqn}) instead is that 
the field $\phi(t,\vec{x})$ can be regarded as {\it classical}. 
Whereas there is little hope of being able to exactly solve the 
Heisenberg field equations for an interacting, four dimensional 
quantum field, {\it one can sometimes solve classical field equations.}
Starobinsky and Yokoyama did it for a minimally coupled scalar with
arbitrary potential \cite{SY}. We have just seen how it can be done
for models with derivative interactions and with constrained fields.
The great task now is to see what happens for quantum gravity.

\vskip .5cm

\centerline{\bf Acknowledgements}

I am grateful to Alexei Starobinski\u{\i} for explaining his technique
and for emphasizing that it recovers the leading infrared logarithms of 
inflationary quantum field theory. This work was partially supported 
by NSF grant PHY-0244714, and by the Institute for Fundamental Theory.

\end{document}